\documentclass[%
 reprint,
 showpacs,preprintnumbers,
 amsmath,amssymb,
 aps,
 prb,
]{revtex4-1}
\usepackage{dcolumn}
\usepackage{subfigure}
\usepackage{here}
\usepackage{graphicx,color}
\usepackage{mathrsfs}

\makeatletter
\def\btt#1{\texttt{\@backslashchar#1}}
\DeclareRobustCommand\bblash{\btt{\@backslashchar}} \makeatother

\begin{document}

\title{Electronic transmission in the lateral heterostructure of semiconducting and metallic transition-metal dichalcogenide monolayers}

\author{Tetsuro Habe}
\affiliation{Department of Applied Physics, Hokkaido University, Spporo, Hokkaido 060-0808, Japan}

\date{\today}

\begin{abstract}
We investigate the electronic transport property of lateral heterojunctions of semiconducting and metallic transition-metal dichalcogenide monolayers, MoSe$_2$ and NbSe$_2$, respectively.
We calculate the electronic transmission probability by using a multi-orbital tight-binding model based on the first-principles band structure. 
The transmission probability depends on the spin and valley degrees of freedom. 
This dependence qualitatively changes by the interface structure.
The heterostructure with a zig-zag interface preserves the spin and the valley of electron in the transmission process. 
On the other hand, the armchair interface enables conduction electrons to transmit with changing the valley and increases the conductance in hole-doped junctions due to the valley-flip transmission. 
We also discuss the spin and valley polarizations of electronic current in the heterojunctions.
\end{abstract}

\maketitle
\section{Introduction}
Transition-metal dichalcogenides (TMDCs) are layered materials of atomically thin two-dimensional crystal consisting of transition-metal and chalcogen atoms, and the monolayer has been attracted much attention in condensed matter physics.
In semiconducting TMDC monolayers, electrons have two discrete degrees of freedom, spin and valley, due to the band structure.\cite{Xiao2012,Zibouche2014}
Strong spin-orbit coupling (SOC) leads to spin-related physics in TMDCs,\cite{Shan2013,Klinovaja2013} and the valley-dependent Berry curvature causes valley Hall effect.\cite{Mak2012}
Therefore, TMDC monolayers have been researched for applications in spintronics\cite{Suzuki2014,Yuan2014,Habe2015,Shao2016} and valleytronics\cite{Cao2012,Zeng2012,Gong2013,Ye2016}.
Since the electronic spin direction is locked to the valley due to the strong SOC, the valleytronics and the spintronics can be combined, e.g., the valley can be detected as the spin.

Stable monolayers can be fabricated by cleaving from a crystal\cite{Helveg2000,Mak2010,Coleman2011} or chemical vapor deposition (CVD) on a substrate\cite{Lee2012,Dong2017}. 
The experimental techniques enable to compose heterostructures consisting of different atomic layers.
The first-designed heterostructure is a stacking of different atomic layers, so-called van der Waals heterostructure, and it has opened several areas of research and application: optics\cite{He2014}, electric transport\cite{Ji2017}, and exciton physics\cite{Baranowski2017,Xu2018}.
The lateral heterostructure is composed of two atomic layers bonded as a single layer, and that of semiconducting TMDCs has been realized by CVD.\cite{Huang2014,Gong2014,Duan2014,Chen2015,Chen2015-2,Zhang2015,He2016}
The semiconducting heterojunction has been investigated theoretically\cite{Kang2013,Habe2015} and experimentally applied to p-n junction\cite{Li2015,Najmzadeh2016}, valleytronics\cite{Ullah2017}, and optoelectronics\cite{Son2016}.
Such the junction provides novel electronic properties for semiconducting TMDCs.

Metallic TMDCs have also attracted much attention in terms of condensed phases.
The TMDC of group-V transition metal atom, Nb and Ta, is metallic and has been discovered to show superconductivity\cite{Lu2015,Wang2017} and the charge density wave (CDW)\cite{Ugeda2015,Xi2015}, even in the monolayer.
NbSe$_2$ monolayer, even in the normal phase, has qualitatively different property of conduction electrons from the semiconducting monolayers.
There are three Fermi pockets in the first Brillouin zone whereas the semiconducting members have two pockets called the $K$ and $K'$ valleys.
Each pocket has both the up-spin and down-spin states.
Thus, the correlation between the spin and valley degrees of freedom, the key property of semicondcuting monolayers for combining the spintronics and the valleytronics , is absent in the metallic monolayers. 
We consider the lateral heterojunction of the metallic and semiconducting TMDC monolayers and show that it induces the correlation between the spin and the three valleys.

\begin{figure}[htbp]
\begin{center}
 \includegraphics[width=55mm]{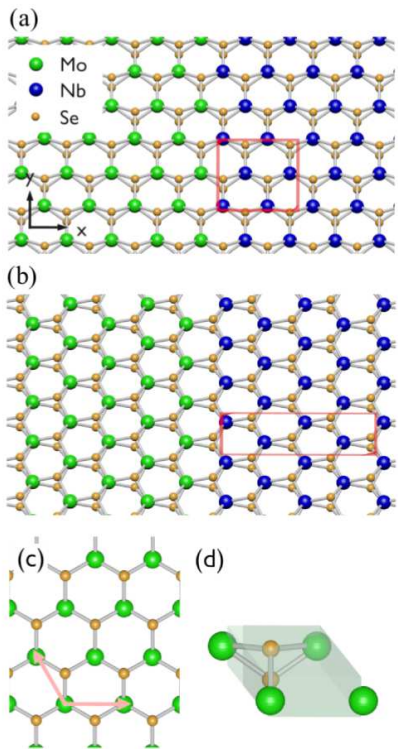}
\caption{
The schematics of atomic heterostructure of MoSe$_2$ and NbSe$_2$ with an armchair interface (a) and a zig-zag interface (b). The unit cell of this junction is indicated by a square frame in (a) and (b). The crystal structure and the primitive unit cell of TMDC monolayer in (c) and (d), respectively. The two arrows indicate the lattice vectors.
 }\label{fig_schematic}
\end{center}
\end{figure}
In this paper, we report our investigation of the spin and valley-dependent electronic transport property of the lateral heterojunction of MoSe$_2$ and NbSe$_2$ monolayers, metallic and semiconducting TMDCs, respectively.
We consider two types of interface for such the junction as shown in Fig.\ \ref{fig_schematic}.
The electronic transmission is qualitatively different in two types of interface, the armchair interface in (a) and zig-zag interface in (b).
We compute the spin and valley-dependence of electronic conductance by using first-principles band calculation and lattice Green's function method, and we show the valley-spin correlated transmission effect.
This result gives a fundamental knowledge for spintronics and valleytronics in lateral heterojunctions.

\section{Band structure}\label{sec_band}
\begin{figure}[htbp]
\begin{center}
 \includegraphics[width=65mm]{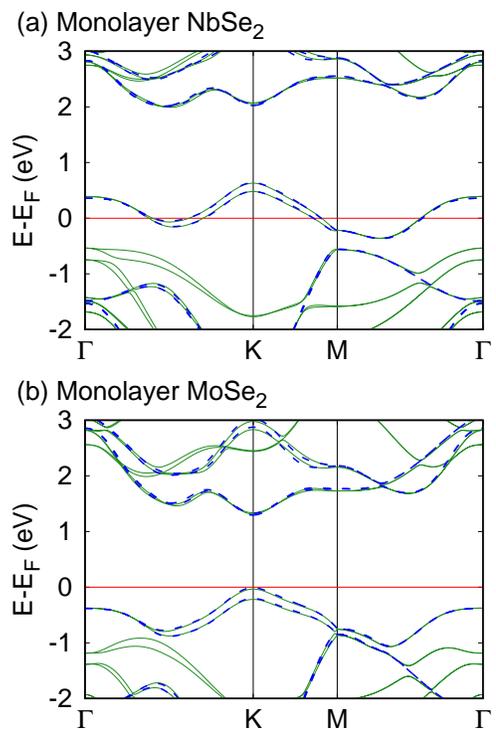}
\caption{
Band structure of monolayer NbSe$_2$ in (a) and MoSe$_2$ in (b). The horizontal line implies the Fermi energy. The dashed lines indicate the bands computed by a multi-orbital tight-binding model (see Sec.\ \ref{sec_tight-binding}).
 }\label{fig_band_structure}
\end{center}
\end{figure}

The TMDC monolayer consists of three sublayers, a sublayer of transition-metal atoms sandwiched by two sublayers of chalcogen atoms.
The transition-metal atoms, Nb or Mo, and chalcogen atoms, Se, are strongly bonded and form two-dimensional hexagonal lattice as shown in Fig.\ \ref{fig_schematic} (c).
We plot the band structure of pristine monolayer of NbSe$_2$ and MoSe$_2$ in Figs.\ \ref{fig_band_structure} (a) and (b), respectively.
These bands are calculated by using quantum-ESPRESSO, a first-principles calculation code\cite{quantum-espresso}. 
Here, we adopt a projector augmented wave (PAW) method with a generalized gradient approximation (GGA) functional including SOC, the cut-off energy of plane wave basis 50 Ry, and the conversion criterion 10$^{-8}$ Ry.
The lattice parameters are also calculated by lattice-relaxation code of quantum-ESPRESSO as $a_{\mathrm{Nb-Nb}}=3.476$\AA\ and $d_{\mathrm{Se-Se}}=3.514$\AA\ in NbSe$_2$, and $a_{\mathrm{Mo-Mo}}=3.319$\AA\ and $d_{\mathrm{Se-Se}}=3.343$\AA\ in MoSe$_2$, where $a_{M-M}$ ($d_{\mathrm{X-X}}$) is the horizontal (vertical) distance between nearest neighbor transition-metal (chalcogen) atoms.

\begin{figure}[htbp]
\begin{center}
 \includegraphics[width=70mm]{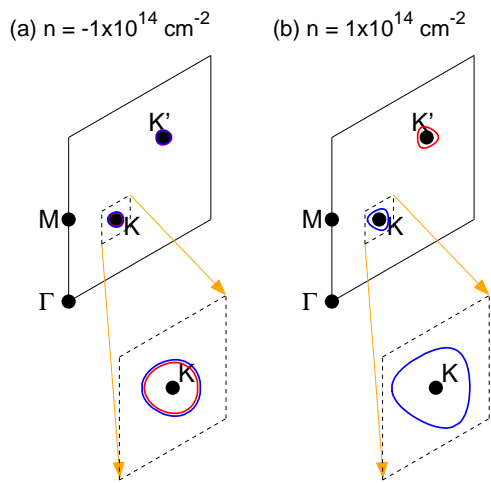}
\caption{
Fermi surface in (a) electron-doped and (b) hole-doped MoSe$_2$. Up-spin and down-spin states are indicated by blue and red, respectively.
 }\label{fig_Fermi_surface_MoSe2}
\end{center}
\end{figure}

MoSe$_2$ monolayer is a semiconductor and has a direct gap at the $K$ and $K'$ points in the Brillouin zone.
At the band edge, the electronic states split into two spin states due to SOC (see Fig.\ \ref{fig_band_structure}), where the spin-split in the valence band is much larger than that in the conduction band.
Here, the spin axis is restricted in the out-of-plane direction due to crystal symmetry.\cite{Xiao2012}
We show the Fermi surface of electron-doped and hole-doped monolayers with the charge density $n=-1\times10^{14}$ cm$^{-2}$ and $1\times10^{14}$ cm$^{-2}$, respectively, in Fig.\ \ref{fig_Fermi_surface_MoSe2}.
The Fermi pockets appear around the $K$ and $K'$ points, and each pocket is so-called a valley.
In the hole-doped monolayer, the spin is fully polarized in each valley up to $|n|\sim10^{14}$ cm$^{-2}$, the experimentally feasible charge density by using electrostatic gating.\cite{Zhang2012-2}
The spin polarization direction is opposite in two valleys due to time-reversal symmetry, and total spin-polarization is absent.
Under the condition of $n<0$, on the other hand, both the spin states are present in each valley.
The electrons in MoSe$_2$ can be characterized by spin and valley indexes at any charge density.

\begin{figure}[htbp]
\begin{center}
 \includegraphics[width=70mm]{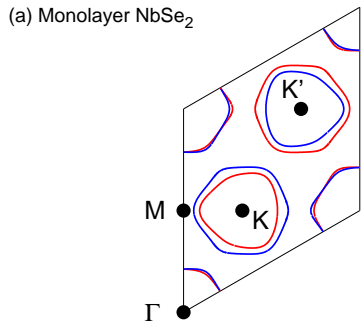}
\caption{
Fermi surface in pristine monolayer NbSe$_2$.
 }\label{fig_Fermi_surface_NbSe2}
\end{center}
\end{figure}
In NbSe$_2$ monolayer, the band is quite similar to that of MoSe$_2$ due to the crystal structure as shown in Fig.\ \ref{fig_band_structure}, but the band is partially filled even at the charge neutral point because the atomic number of Nb is one less than Mo.
We show the Fermi surface of pristine NbSe$_2$ monolayer in Fig.\ \ref{fig_Fermi_surface_NbSe2}.
The Fermi pockets appear around three high-symmetry points, the $K$, $K'$, and $\Gamma$ points. 
We represent the pockets as the $K$, $K'$, and $\Gamma$ valleys in what follows.
In every valleys, both the two spin states are present. 
The spin polarization is non-zero and opposite each other in the $K$ and $K'$ valleys, but the populations of two spins are equal to each other in the $\Gamma$ valley.
Every Fermi pockets are strongly trigonal warping due to three-fold symmetry of crystal structure.

\section{Calculation method}\label{sec_tight-binding}
To calculate the electronic transmission probability, we adopt a tight-binding model in bases of Wannier functions, where the hopping integrals are computed from the first-principles band structure in Sec.\ \ref{sec_band}.
The maximally localized Wanner functions and spin-dependent hopping integrals are computed by using Wannier90\cite{wannier90}.
Here, we adopt three $d$-orbitals $|M,d_{3z^2-r^2}\rangle$, $|M,d_{xy}\rangle$, and $|M,d_{x^2-y^2}\rangle$ of transition-metal atom $M$, and three superpositions of $p$-orbitals of top selenium Se$^{t}$ and bottom one Se$^{b}$ as $|\mathrm{Se}^+,p_{x}\rangle$, $|\mathrm{Se}^+,p_{y}\rangle$, and $|\mathrm{Se}^-,p_{z}\rangle$ with $|\mathrm{Se}^\pm,p_{\nu}\rangle\equiv(|\mathrm{Se}^t,p_{\nu}\rangle\pm|\mathrm{Se}^b,p_{\nu}\rangle)/\sqrt{2}$.
These orbitals have even parity under mirror operation in the $z$ axis and they are independent of odd parity orbitals in the electronic structure.
We show the band structure calculated by the tight-binding model in Fig.\ \ref{fig_band_structure}.
It well reproduces the first-principles bands.

We simulate the electronic transmission in lateral heterostructures at zero temperature by using the multi-orbital tight-binding model including the interface between MoSe$_2$ and NbSe$_2$ monolayers.
Here, NbSe$_2$ monolayer shows the phase transition, CDW and superconductivity, at low temperature but we consider normal states of NbSe$_2$ for investigating the fundamental transport property of metallic NbSe$_2$ monolayer.
We assume a commensurate interface, where two atomic layers are bonded without dangling and misalignment as shown in Fig.\ \ref{fig_schematic}, and a periodic boundary condition parallel to the interface.
The incident electronic wave with $k_y$ in MoSe$_2$ transmits to NbSe$_2$ as a wave with $k_y'=(a_{\mathrm{Mo-Mo}}/a_{\mathrm{Nb-Nb}})k_y$, and thus the transmission coefficient is defined at each $k_y$.
In what follows, we represent both the wave numbers, $k_y$ and $k_y'$, by $k_y$. 
Despite the mismatch of lattice parameters, such a commensurate heterostructure has been fabricated experimentally.\cite{Li2015,Gong2014}
To use lattice Green's function method, we consider a unit cell including four transition-metal atoms and eight Se as shown in Fig.\ \ref{fig_schematic}.
Here, we adopt the hopping matrix in NbSe$_2$ as that between MoSe$_2$ and NbSe$_2$ because the outer shell of Mo and Nb atoms is same as 4$d$-orbitals, and the hopping integral is similar to each other in the two monolayers.

We calculate the transmission probability $\tilde{t}_{\mu\nu}(k_y)$ by using lattice Green's function method for multi-orbital tight-binding model\cite{Ando1991,Habe2015,Habe2016}, where $\nu$ and $\mu$ represent the incoming and outgoing waves in MoSe$_2$ and NbSe$_2$, respectively. 
The conductance per unit width is given by 
\begin{align}
g_e=\frac{e^2}{\pi\hbar}\int \frac{dk_y}{2\pi}\sum_{\mu,\nu}\left|\sqrt{\frac{v_\mu}{v_\nu}}\tilde{t}_{\mu\nu}(k_y)\right|^2.\label{eq_conductance}
\end{align}
Here, $v_\mu$ and $v_\nu$ are the velocity along the $x$ axis for the outgoing wave in NbSe$_2$ and the incoming wave in MoSe$_2$, respecitvely.
We define the positive direction in the $x$ axis as the transmission direction.
Thus, the electronic waves $\mu$ and $\nu$ have the positive velocity in the $x$ axis.
The wave function is represented by the periodic part $\boldsymbol{c}_j$, wave amplitudes of Wanner functions, in a unit cell and the phase shift $\lambda=e^{-i\phi}$ in the transmission process between adjacent cells. 
The coefficients are the eigenfunction and eigenvalue of the following equation for each $k_y$ in pristine MoSe$_2$ and NbSe$_2$ monolayers as
\begin{widetext}
\begin{align}
\lambda\begin{pmatrix}
\boldsymbol{c}_j\\
\boldsymbol{c}_{j-1}
\end{pmatrix}
=
\begin{pmatrix}
\hat{h}_1^{-1}(k_y)(\hat{h}_0(k_y)-E_F)&-\hat{h}_1^{-1}(k_y)\hat{h}_{-1}(k_y)\\
1&0
\end{pmatrix}
\begin{pmatrix}
\boldsymbol{c}_j\\
\boldsymbol{c}_{j-1}
\end{pmatrix},
\end{align}
\end{widetext}
with the onsite potential $\hat{h}_0(k_y)$, and the right-going and left-going hopping matrix $\hat{h}_1$ and $\hat{h}_{-1}$, respectively.
The velocity in the $x$ axis is calculated by $v_\mu=(\boldsymbol{c}_j,\boldsymbol{c}_{j-1})^\dagger\hat{v}_x(\boldsymbol{c}_j,\boldsymbol{c}_{j-1})$ with the velocity operator as
\begin{align}
\hat{v}_x=-i\frac{a_x}{\hbar}\begin{pmatrix}
0&\hat{h}_{-1}(k_y)\\
\hat{h}_{1}(k_y)&0
\end{pmatrix},
\end{align}
where $a_x$ is the length of square unit cell in Fig.\ \ref{fig_schematic} along $x$.
Here, the Fermi energy $E_F$ is computed under the condition of a charge density $n$ as
\begin{align}
n=n_0-\int \frac{d^2\boldsymbol{k}}{(2\pi)^2} \theta(E_{\alpha,\boldsymbol{k}}-E_F),
\end{align}
where $n_0$ is the charge density of nuclei, and $E_{\alpha,\boldsymbol{k}}$ is the energy dispersion in Fig.\ \ref{fig_band_structure}.
We consider the charge induction by using a top or back gate being homogeneous in the $xy$ plane.
By gating, opposite charges are induced in the substrate and homogeneously distributed in the $xy$ plane.
Thus, we can assume that the induced charges in the heterojunction are distributed homogeneously, i.e., $n=\mathrm{const.}$, in the $xy$ plane due to the local charge neutrality.
Then, the Fermi energy is also applied to the MoSe$_2$ and NbSe$_2$ regions in the lateral heterojunction.
We also calculate $k_x$ of incoming and outgoing waves by using $\phi=k_xc_x$, and thus we can characterize these waves by the valley degree of freedom $\tau$ by referring to the Fermi surface in Figs.\ \ref{fig_Fermi_surface_MoSe2} and \ref{fig_Fermi_surface_NbSe2}.
Therefore, the wave index, $\mu$ and $\nu$, is given by the valley index $\tau$ and the spin index $s$.  
By picking up a spin and a valley, it is possible to calculate the conductance of spin-polarized electrons transferring between two valleys.
\begin{figure}[htbp]
\begin{center}
 \includegraphics[width=55mm]{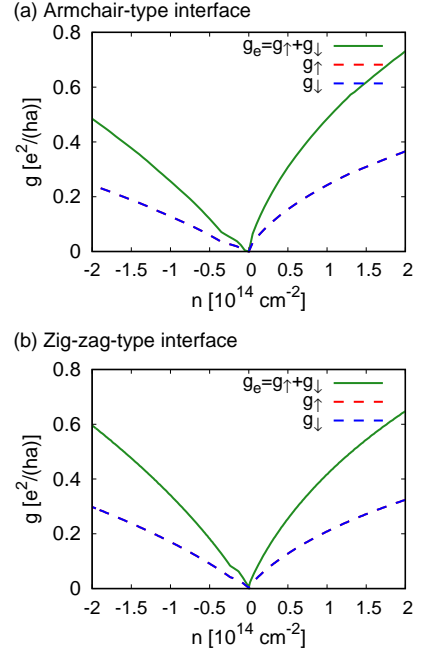}
\caption{
The charge density-dependence of electric conductance per unit width (solid line) in the heterojunction with an armchair-type interface (a) and a zig-zag-type interface (b). The dashed lines indicate the conductivity for each spin state.
 }\label{fig_conductivity}
\end{center}
\end{figure}
\begin{figure}[htbp]
\begin{center}
 \includegraphics[width=55mm]{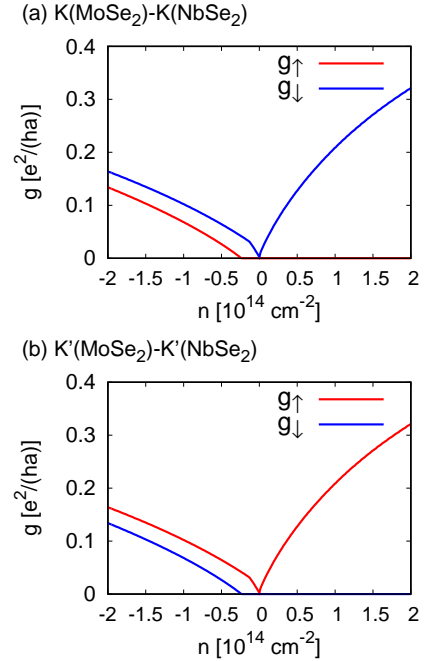}
\caption{
Valley and spin dependence of electric conductance in lateral heterostructure with a zig-zag interface. (a) and (b) represent the transmission of electron in the $K$ and $K'$ valleys, respectively. 
 }\label{fig_zig-zag}
\end{center}
\end{figure}

\section{Numerical results}
First, we show the charge density-dependence of electric conductance $g_e$ in Fig.\ \ref{fig_conductivity}.
The conductance drops around $n=0$ in both the two types of heterostructures with an armchair-type interface and a zig-zag-type interface in Fig.\ \ref{fig_schematic} (a) and (b), respectively, and it monotonically increases with $|n|$.
This is because the Fermi level is in the insulating gap under $n=0$, and the conducting channels increase with $|n|$ in MoSe$_2$.
At every non-zero charge densities, the heterostructure with the armchair-type interface gives a larger $g_e$ than that with the zig-zag-type interface.
The armchair interface allows electrons transferring between different valleys, e.g., $K$ and $\Gamma$, in MoSe$_2$ and NbSe$_2$ as discussed below, and thus it enhances the total transmission probability.  
In the transmission process, the electronic spin is conserved due to the mirror symmetry in the $z$ axis, the out-of-plane direction.
Thus, the conductance $g_e$ is separated into that of up-spin electrons $g_\uparrow$ and down-spin electrons $g_\downarrow$.
In TMDC monolayers, the spin relaxation is suppressed even in the presence of non-magnetic impurity due to the mirror symmetry.\cite{Ochoa2013-1,Ochoa2013-2,Habe2016} 
The two spin components are balanced due to the equal population of two spins in the presence of time-reversal symmetry.

\begin{figure*}[htbp]
\begin{center}
 \includegraphics[width=160mm]{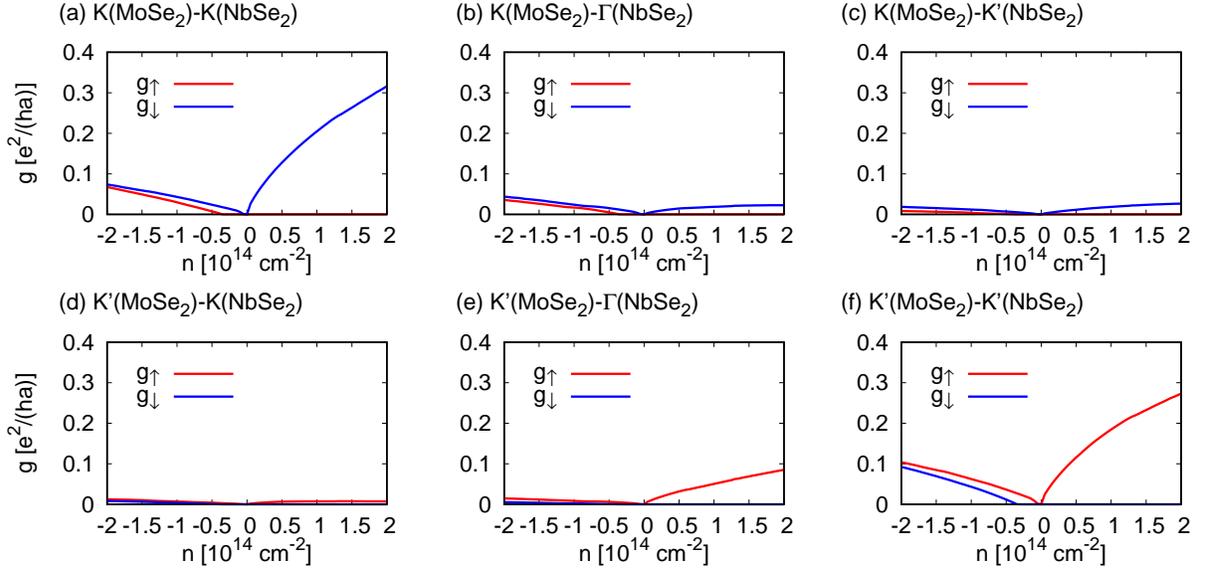}
\caption{
Valley and spin dependence of electric conductance in lateral heterostructure with an armchair interface. Upper panels (a)-(c) and lower panels (d)-(f) represent the transmission of electron in the $K$ and $K'$ valleys of MoSe$_2$.
 }\label{fig_armchair}
\end{center}
\end{figure*}
Nest, we show the valley-dependent conductance in the two types of heterostructures in Figs.\ \ref{fig_zig-zag} and \ref{fig_armchair}.
The heterostructure with a zig-zag interface preserves mirror symmetry in the $y$ axis.
Thus, the transmission probability is equivalent in both the valleys as shown in Fig.\ \ref{fig_zig-zag}.
When passing through the zig-zag interface, electrons preserve the valley degree of freedom due to conservation of wave number $k_y$ parallel to the interface.
Here, $k_y$ is parallel to the $M-K$ direction in Figs.\ \ref{fig_Fermi_surface_MoSe2} and \ref{fig_Fermi_surface_NbSe2}, and the transmission of electron with $k_y$ is non-zero as long as electronic states are present in the Fermi surface of both MoSe$_2$ and NbSe$_2$.
Since the spin is fully polarized in the $K$ and $K'$ valleys on the condition of $0<n$ in MoSe$_2$, the spin and valley are locked to each other in the transmission process.
In $n<0$, on the other hand, the up-spin and down-spin electrons flow above $n=3\times10^{13}$ cm$^{-2}$ in both the two valleys.

In Fig.\ \ref{fig_armchair}, we show the spin and valley dependence of conductance of the heterojunction with an armchair-type interface.
Valley-conserving transmission probability is similar to that in the zig-zag interface.
Moreover, electrons in MoSe$_2$ are able to transmit with switching the valley to $\Gamma$ as shown in (b) and (e), but the valley-switch between the $K$ and $K'$ valleys is highly suppressed even in this heterostructure as shown in (c) and (d).
The electronic flow in the $\Gamma$ valley carries the spin current in $0<n$.
\begin{figure}[htbp]
\begin{center}
 \includegraphics[width=70mm]{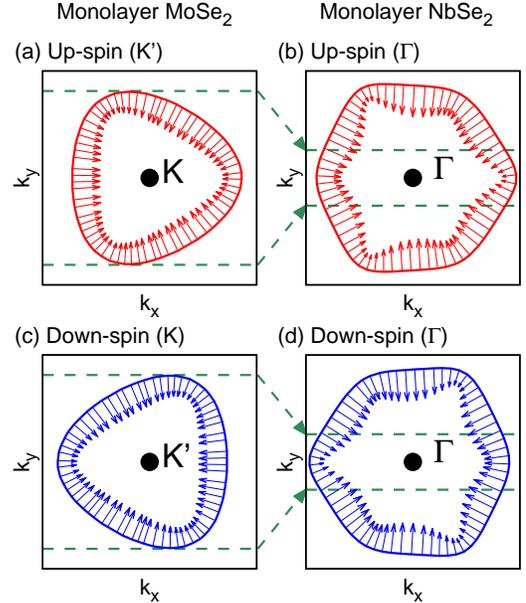}
\caption{
Fermi velocity in MoSe$_2$ and NbSe$_2$ with $n=1\times10^{14}$ [cm$^{-2}$]. Dashed lines indicate the upper and lower $k_y$ at which conducting channels present in MoSe$_2$. The arrow represents the Fermi velocity, and its length is the relative speed.
 }\label{fig_velocity}
\end{center}
\end{figure}
The spin flow is attributed to the difference of transmission coefficient for incoming electronic waves in the $K$ and $K'$ valleys of MoSe$_2$.
Here, the transmission coefficient is defined as $t_{\mu\nu}=\sqrt{v_\mu/v_\nu}\tilde{t}_{\mu\nu}$, which is the integrand in Eq.\ (\ref{eq_conductance}), and it depends on the ratio of Fermi velocity $v_\mu/v_\nu$ in the pristine NbSe$_2$ and MoSe$_2$.
We show the Fermi velocity in MoSe$_2$ and NbSe$_2$ in Fig.\ \ref{fig_velocity}, where the electronic states having the right-going velocity and being inside of dashed lines are corresponding to the wave vector holding conducting channels.
The Fermi velocity of right-going states with down-spin is much smaller than that with up-spin in NbSe$_2$.
The spin current in the $\Gamma$ valley is caused by this anisotropy of Fermi velocity.
The armchair interface allows us to obtain not only the spin-polarized electronic current in the $\Gamma$ valleys but also the valley polarized current between the $K$ and $K'$ valleys, which can be observed by valley Hall effect due to the opposite Berry curvature in the two valleys.\cite{Xiao2012,Shan2013,Habe2017}

\section{Discussion}
\begin{figure}[htbp]
\begin{center}
 \includegraphics[width=65mm]{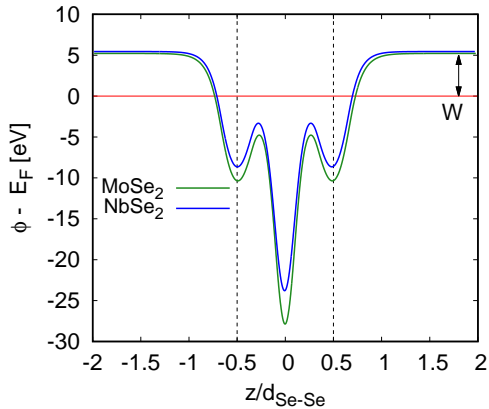}
\caption{
Spatially electrostatic potential as a function of the out-of-plane coordinate in pristine MoSe$_2$ and NbSe$_2$ monolayers.
The dashed lines indicate the positions of chalcogen atoms.
 }\label{fig_work_function}
\end{center}
\end{figure}
In the previous section, we consider the heterojunction with a charge density induced by electrostatic gating.
In this section, we discuss the effect of chemical doping in the lateral heterojunction.
The dopants form impurity levels inside the gap of semiconductor and pin the Fermi energy at a level without gating.
Around the interface, the charge density is modulated due to the difference of work function between MoSe$_2$ and NbSe$_2$.
We calculate the electrostatic potential for electrons as a function of $z$, the out-of-plane coordinate, by using quantum-ESPRESSO and plot it referring to the Fermi level in Fig.\ \ref{fig_work_function}.
Here, we set the origin of $z$ at the transition-metal atoms and adopt the potential at $z/d_{\mathrm{Se}-\mathrm{Se}}=2$ as the vacuum level.
The work function is defined by the vacuum level with respect to the Fermi energy.
We compute the work function $W_{\mathrm{MoSe}_2}=5.217$ eV and $W_{\mathrm{NbSe}_2}=5.447$ eV, where $W_{\mathrm{MoSe}_2}$ is similar to a previous work\cite{Lanzillo2015}, and obtain the difference $W_{\mathrm{MoSe}_2}-W_{\mathrm{NbSe}_2}=-0.230$ eV.
Since the semiconducting TMDC, MoSe$_2$, has a smaller work function than the metallic TMDC, NbSe$_2$, a Schottky barrier is formed in the junction of electron-doped MoSe$_2$ due to the modulation of doped carrier density.
In the hole-doped heterostructure, on the other hand, the junction has a Ohmic contact.

Around the interface of realistic materials, charges transfer between two TMDCs due to the difference of work functions. 
The charge transfer modifies the Fermi energy near the interface and it can be treated as a contact resistance. 
It reduces the transmission probability but the spin and valley polarization is unchanged qualitatively because time-reversal and mirror symmetries are preserved.
Furthermore, the control of contact resistance for atomic layered materials has been developing, e.g., graphene-metal junction\cite{Giubileo2017}.
In our calculation, we assume the uniform Fermi energy over each TMDC for analyzing an ideal transmission case.

\begin{figure}[htbp]
\begin{center}
 \includegraphics[width=65mm]{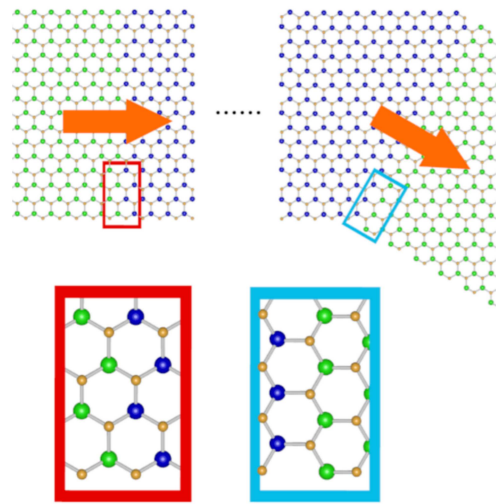}
\caption{
The schematic of MoSe$2$-NbSe$_2$-MoSe$_2$ junction with two boundaries. The arrows indicate the current direction.
 }\label{fig_app}
\end{center}
\end{figure}

Finally, we discuss the experimental observation associated with the spin and valley dependent transmission in the lateral heterojunction.
We show that the spin of conduction electrons is not polarized at a single heterojunction as shown in Fig.\ \ref{fig_conductivity}. 
By combining the armchair and zig-zag interfaces, on the other hand, the two-interface heterojunction induces the spin and valley polarized current.
In Fig.\ \ref{fig_app}, the left interface changes the transmission probability between the $K$ and $K'$, and the right interface allows the equal transmission for the two valleys.
Then, the spin and valley polarization induced by the left interface remains in the right MoSe$_2$ region.
Therefore, the spin and valley dependent transmission can be shown in the two-interface hererojunction experimentally.

\section{Conclusion}
We have investigated the electronic transport property of lateral heterojunctions of metallic NbSe$_2$ monolayer and semiconducting MoSe$_2$ monolayer and shown that the spin and valley dependence of conductance changes with the interface structure.
The zig-zag interface conserves the spin and valley of conduction electron in the $K$ and $K'$ valleys, and the conductance is independent of the valley and spin.
This means that NbSe$_2$ can be applied to the lead in spintronic or valleytronic devices of semiconducting TMDCs.
When electrons pass through an armchair interface, the electronic spin is also conserved, but the valley, on the other hand, changes in the transmission process. 
This valley-flip increases the total transmission probability, and thus the conductance is larger than that with the zig-zag interface in $0<n$.
Moreover, the transmission probability is strongly depending on the spin and valley, and thus conduction electrons are spin-polarized in each valley.
The lateral heterostructure can be useful in spintronics and valleytronics, and its transport properties studied here provide the opportunity to develop devices made by only atomic layers.

\bibliography{TMDC}
\end{document}